\documentclass{article}

\usepackage{graphicx} 
\usepackage[papersize={8.5in,11in}, margin=1in]{geometry}
\usepackage{setspace}
\usepackage{xcolor}
\usepackage{amsmath}
\usepackage{caption}
\usepackage{ragged2e}
\usepackage{authblk}
\usepackage[colorlinks=true,citecolor=black,linkcolor=black]{hyperref}
\usepackage[superscript,biblabel]{cite} 
\usepackage{caption}
\usepackage{subfiles}
\justifying

\title{Linear-Scaling Potential-Free Data-Driven Molecular Dynamics for Arbitrary-Sized Water Clusters (H\textsubscript{2}O)\textsubscript{n}}

\author[1]{Hongyu Yan}
\author[2*]{Yong Wei}
\author[3*]{Minghan Chen}
\author[4*]{Hanning Chen}

\affil[1]{School of Engineering and Physical Sciences, Heriot-Watt University, Edinburgh EH14 4AS, UK}
\affil[2]{Department of Computer Science and Cybersecurity, University of North Georgia, Dahlonega, Georgia 30597, USA}
\affil[3]{Department of Computer Science, Wake Forest University, Winston-Salem, North Carolina 27104, USA}
\affil[4]{Texas Advanced Computing Center, The University of Texas at Austin, Austin, Texas 78758, USA}

\affil[*]{Corresponding authors: $^2$yong.wei@ung.edu; $^3$chenm@wfu.edu; $^4$hanning.chen@austin.utexas.edu}

\date{}

\begin{document}

\maketitle

\begin{abstract}
\doublespacing
Conventional molecular dynamics (MD) simulation approaches, such as \textit{ab initio} MD (AIMD) and empirical force field MD (EFFMD), face significant trade-offs between physical accuracy and computational efficiency. This work presents a linear-scaling potential-free data-driven molecular dynamics (PDMD) framework for predicting system energy and atomic forces of arbitrary-sized water clusters (H\textsubscript{2}O)\textsubscript{n}. Specifically,
PDMD employs a Gaussian-based geometry descriptor to generate high-dimensional, atomistic footprints,  then leverages ChemGNN, a graph neural network model that adaptively learns the atomic chemical environments without requiring \textit{a priori} knowledge.
Through an iterative self-consistent training approach, the converged PDMD achieves a mean absolute error of 1.39 meV/atom for energy, outperforming other state-of-the-art models such as DeepMD, MACE, NequIP, and SevenNet by at least 2.6x in accuracy with the same dataset. As a result, the linear-scaling PDMD can reproduce the AIMD properties of water clusters at orders-of-magnitude lower computational cost, as illustrated by simulations of systems consisting of thousands or more molecules. These results demonstrate that the proposed PDMD offers multiphase predictive power and enables ultra-fast, general-purpose MD simulations while retaining AIMD-level accuracy. This accuracy is achieved by efficiently capturing many-body potentials that are critical in numerous polyatomic systems but are often missing in EFFMD. Moreover, we have constructed an \textit{ab initio} dataset with over 300,000 (H\textsubscript{2}O)\textsubscript{n} structures, standardized in a unified PyTorch Geometric framework, to support scalable evaluation of artificial intelligence methods for molecular dynamics.

\end{abstract}

\setstretch{2}

\section*{Introduction}
Molecular dynamics (MD) is one of the most popular numerical techniques for simulating the temporal and spatial evolution of chemical and biological systems.\cite{karplus2002molecular} The quality of an MD simulation rests on the accuracy of its system energy, \textit{E}, and atomic forces, $\{\overrightarrow{F_i}\}_{i=1...N_{atom}}$, typically evaluated as a function of atom types and atomic coordinates,\cite{mouvet2022recent} possibly in addition to bonding topology\cite{lifson1968consistent} and atomic polarizability.\cite{warshel2007polarizable} For instance, in an \textit{ab initio} MD (AIMD) simulation,\cite{iftimie2005ab} \textit{E} is obtained by diagonalizing a quantum Hamiltonian matrix while $\{\overrightarrow{F_i}\}$ is given by the Hellmann-Feynman theorem.\cite{feynman1939forces} In general, AIMD simulations are computationally demanding as the time to resolve the many-body electron wavefunction, $\mathit{\Psi}$, proliferates over electronic degrees of freedom, $N_{elec}$. Even under the mean-field approximation for single-particle orbitals such as in density functional theory (DFT),\cite{hohenberg1964inhomogeneous} AIMD's computational complexity is proportional to $O(N_{elec}^3)$, limiting its application to large-scale systems. 

A numerically efficient alternative to AIMD is the empirical force field MD (EFFMD), wherein \textit{E} is expressed as the sum of many-body bonded energies and pairwise nonbonded potentials for a given bonding topology. In modern empirical force fields (EFFs), such as CHARMM,\cite{best2012optimization} AMBER,\cite{tian2019ff19sb} GROMOS,\cite{reif2012new} and OPLS,\cite{harder2016opls3} the bonded terms up to four-body are included to reflect bond stretching, angle bending, and dihedral torsion. By contrast, the nonbonded potentials consist of Coulomb coupling and van der Waals (vdW) forces, which can be supplemented by electric polarization to explicitly account for dispersion and induction effects such as in AMOEBA.\cite{shi2013polarizable} Usually, EFFs are parameterized against quantum potential energy surfaces (PES)\cite{huang2013automated} and/or experimental data\cite{frohlking2020toward} at equilibrium states, making them unsuitable for systems that are far from equilibrium. For example, the bond stretching potential, \textit{E\textsubscript{b}}, between two atoms commonly adopts a shifted quadratic function centered at their equilibrium bond length, \textit{r\textsubscript{0}}, i.e., $E_{b}=\frac{1}{2}k(r-r_{0})^2$. This harmonic approximation is mostly reasonable for the small oscillations of a chemical bond, whereas it becomes invalid upon bond dissociation due to the increasing anharmonicity of PES. Aiming to remedy the oversimplified interatomic potentials in conventional EFFs, reactive force fields (RFFs) such as ReaxFF\cite{senftle2016reaxff} and MS-EVB\cite{wu2008improved} were developed to allow for more flexible potential functions and more dynamical bonding topology, enabling the simulations of chemical reactions. Nevertheless, a fully-fledged RFF demands an exhaustive exploration of the parameter space to fully reproduce a quantum PES, the majority of which is often unknown for condensed-phase systems. Moreover, the escalated complexity of RFF significantly increases its computer execution time compared to EFF, largely eroding its efficiency advantage over AIMD. As a result, a novel approach to attain \textit{E} and $\{\overrightarrow{F_i}\}$ with AIMD's level of accuracy but at the computational cost of EFF is highly desired.

Over the past decade, artificial intelligence (AI) has become a powerful tool for molecular simulations\cite{noe2020machine} thanks to the rapid advances in specialized hardware acceleration and machine learning (ML) algorithms. The greatest advantage of data-driven AI over traditional physics-based simulations is its independence from the governing physical equations,\cite{montans2019data} whose solutions are often too difficult or nearly impossible to compute. For example, the exact solution of a many-electron wavefunction, $\mathit{\Psi}$, in its ground state is unavailable due to electron correlation,\cite{raghavachari1991electron} which, however, can be numerically recovered by expressing $\mathit{\Psi}$ as a weighted linear combination of Slater determinants of one-electron orbitals, $\{\phi_{n}\}$, i.e.,$\mathit{\Psi=\sum_{i}}c_{i}det[\{\phi_{n}\}^i]$, where the weights, $\{\mathit{c_{i}}\}$, are optimized by quantum Monte Carlo (QMC)\cite{mcmillan1965ground} or full configuration interaction (FCI)\cite{ross1952calculations}.  In theory, countless Slater determinants could be included, but ML approaches such as the Fermionic Neural Network (FNN)\cite{pfau2020ab} filters out those with high energies and thus negligible weights. Using atomic positions $\overrightarrow{X_i}$ and atom types $Z_i$ as inputs, the FNN makes QMC simulations feasible beyond simple models.

Machine learning has been utilized to accelerate computational simulations by reducing the computational cost of complex models. In fluid dynamics, a convolutional neural network (CNN)\cite{kochkov2021machine} has been employed to accelerate the resolution of time-dependent velocity fields by orders of magnitude without prior knowledge of the Navier-Stokes equations\cite{glowinski1992finite}. More recently, a generative adversarial network (GAN)\cite{li2021ai} empowered the high-resolution simulation of the universe's evolution over its full dynamic range using only low-resolution astronomical images collected from the Hubble Space Telescope. Similarly, many efforts have been devoted to developing machine learning force fields (MLFF),\cite{unke2021machine} hoping to reproduce high-quality \textit{ab initio} PES from a minimal set of chemical descriptors such as atom types, interatomic distances, and partial charges. Unlike the conventional kernel-based methods\cite{scholkopf1998nonlinear} that map the input chemical descriptors to a higher-dimensional feature space for linear regression, MLFFs, inspired by neural networks\cite{fukushima1980neocognitron}, decompose the nonlinear dependency of \textit{ab initio} PES into a linear combination of learnable parameters connected by nonlinear activation functions.\cite{fukushima1969visual} For example, two fully connected hidden layers, each featuring 25 nodes, accurately predicted liquid water's relatively weak yet physically essential vdW potential compared to the BLYP density functional\cite{becke1988density} after BP-NNP\cite{morawietz2016van} transforms type 2 and type 4 interatomic distance-dependent symmetry functions\cite{behler2011atom} into an input vector for the dense neural network. Since BP-NNP\cite{morawietz2016van} and other DNN- or CNN-based MLFFs such as ANI\cite{smith2017ani} and SchNet\cite{schutt2017schnet} cannot encode bonding topology, they often fail to distinguish bonded and non-bonded interactions, which differ remarkably in strength, range, and form. 

Graph neural networks (GNNs)\cite{scarselli2008graph} have emerged as a more natural representation of chemical systems in ML. In GNN-inspired MLFFs, the atoms and bonds are represented as nodes and edges, respectively, in a graph wherein the embedding at each node is updated by its neighbors' message functions.\cite{hasebe2021knowledge}. For example, in the graph neural network force field (GNNFF) model\cite{park2021accurate}, the total force on an atom is assumed to be the sum of the contributions from a fixed number of its neighboring atoms, ensuring the sparsity of the adjacency matrix for the linearity of computational cost with system size. Unfortunately, the missing capability of energy prediction in GNNFF prevents its application under all thermodynamic conditions except for the microcanonical ensemble.\cite{park2021accurate} As an intuitive remedy, the atomistic line graph neural network-based force field (ALIGNN-FF)\cite{choudhary2023unified} adopts a composite loss function consisting of error terms for both forces and energies. However, the relatively low quality of ALIGNN-FF's pairwise potential could artificially destabilize an otherwise well-equilibrated system as evidenced by its large mean absolute error of energy, \textit{E\textsuperscript{MAE}},\cite{choudhary2023unified} relative to the small thermal fluctuation energy, $k_BT$, at room temperature. In a satisfactory scenario for room-temperature MD simulation, \textit{E\textsuperscript{MAE}} is expected to be less than 30 meV/atom.

Chemical Environment Graph Neural Network (ChemGNN) \cite{chen2023chemical} is a new graph neural network model meticulously designed to delve into the intricate dynamics of interatomic interactions within atoms' local chemical environments. The model showed salient performance in predicting the band gaps of graphitic carbon nitride (g-C\textsubscript{3}N\textsubscript{4}) and its doped variants. At the core of its architecture, initial node embeddings and edge features encode the chemical and physical attributes of atoms and their interactions, capturing the essence of their spatial arrangement. Within the framework of ChemGNN, the influence of interatomic interactions is distilled through the aggregation of messages exchanged among neighboring atoms. The model uses a set of permutation-invariant aggregation functions that effectively capture various aspects of these interactions, revealing the underlying structure of chemical systems. Aiming to discern the significance of various aggregated impacts on an atom, each aggregation function is assigned a learnable weight. Consequently, node embeddings are iteratively updated based on the amalgamated messages derived from the local chemical environment of the atoms during model training. These refined node embeddings are then used to make predictions at both the node and graph levels, such as forecasting atomic forces and system energy, thus facilitating a comprehensive understanding of molecular behaviors. 

The unique multiphase predictive power of ChemGNN can be well demonstrated on arbitrary-sized water clusters, (H\textsubscript{2}O)\textsubscript{n}, which witness the onset of a gas-to-liquid phase transition upon adding water molecules, one at a time.\cite{keutsch2001water} As shown by a vibrational spectroscopic study\cite{rognoni2021many}, (H\textsubscript{2}O)\textsubscript{21} exhibits a pronounced red shift of two O-H stretching normal modes by more than 400 cm\textsuperscript{-1}, which was not observed in any smaller water clusters. Moreover, similar to liquid water's ``fingerprint" band\cite{brubach2005signatures}, (H\textsubscript{2}O)\textsubscript{21}'s two red-shifted infrared peaks near 3300 cm\textsuperscript{-1} are relatively broad, confirming (H\textsubscript{2}O)\textsubscript{21} as the minimum network to fully solvate a water molecule at its center. Additionally, this phase transition is driven by the delicate balance between bulk energy gain and surface-tension penalty, according to classical nucleation theory,\cite{jun2022classical} which predicts a minimum critical size of $\sim$2.0 nm (i.e., $\sim$350 H\textsubscript{2}O molecules) for water nucleation, a prediction confirmed by a microemulsion study under supercooling.\cite{liu2007direct} Interestingly, the critical size for water nucleation depends strongly on the degree of supersaturation, leading to distinct condensation rates for cloud formation.\cite{charlson2001reshaping} For example, under typical atmospheric conditions with $\sim0.2\%$ supersaturation, the critical size is $\sim60$ nm, approximately twice that at a higher supersaturation of $\sim0.5\%$ in clean air.\cite{svensmark2024supersaturation} Nevertheless, even a water cluster with a critical size of $\sim30$ nm has more than 1 million atoms, which is well beyond the computational capacity of any \textit{ab initio} molecular dynamics simulation today, making the high-fidelity modeling of microscopic water clusters a paramount challenge. Therefore, the (H\textsubscript{2}O)\textsubscript{n} clusters that offer a diverse set of multi-phase chemical environments were chosen as our test bed to assess the performance of ChemGNN, which is the core of our potential-free data-driven molecular dynamics (PDMD) by affording $\textit{E}$ and ${\overrightarrow{F_i}}$ out of $\overrightarrow{X_i}$ and $Z_i$. Most importantly, our PDMD framework presented in this work for arbitrary-sized (H\textsubscript{2}O)\textsubscript{n} clusters in the gas phase could be readily extended to condensed-phase systems, such as aqueous solutions, the primary medium for numerous chemical reactions and most biological processes.   

\section*{Results}

\subsection*{a. Performance of PDMD Model on Energy and Force Prediction}

\begin{figure}[!htbp]
\centering
\includegraphics[width=0.99\textwidth]{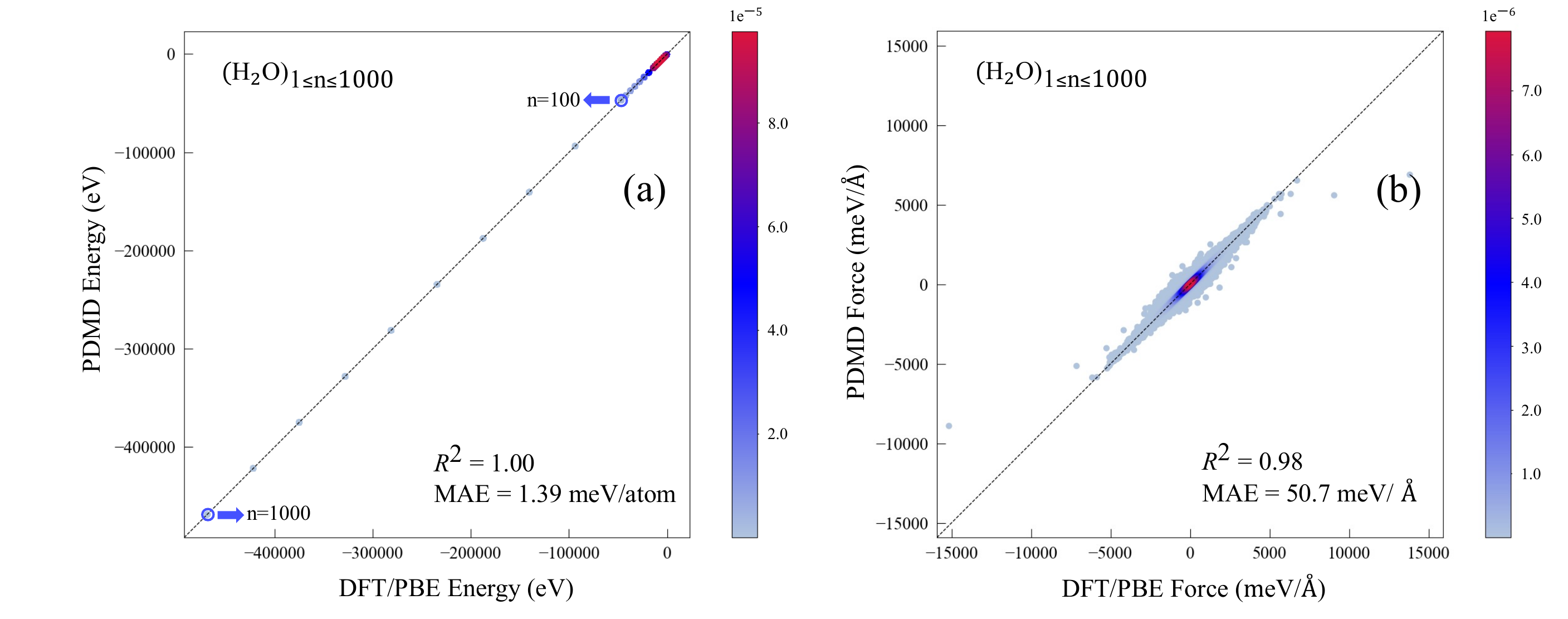}
\caption{The mean absolute errors (MAEs) of PDMD for (a) system energy, $\textit{E}$, and (b) atomic force, ${\overrightarrow{F_i}}$, across all (H\textsubscript{2}O)\textsubscript{1$\leq$n$\leq$1000} clusters. The data point density is reflected by color bars.}
\label{fig:acc_ChemGNN}
\end{figure}

\begin{figure}[!t]
\centering
\includegraphics[width=0.6\textwidth]{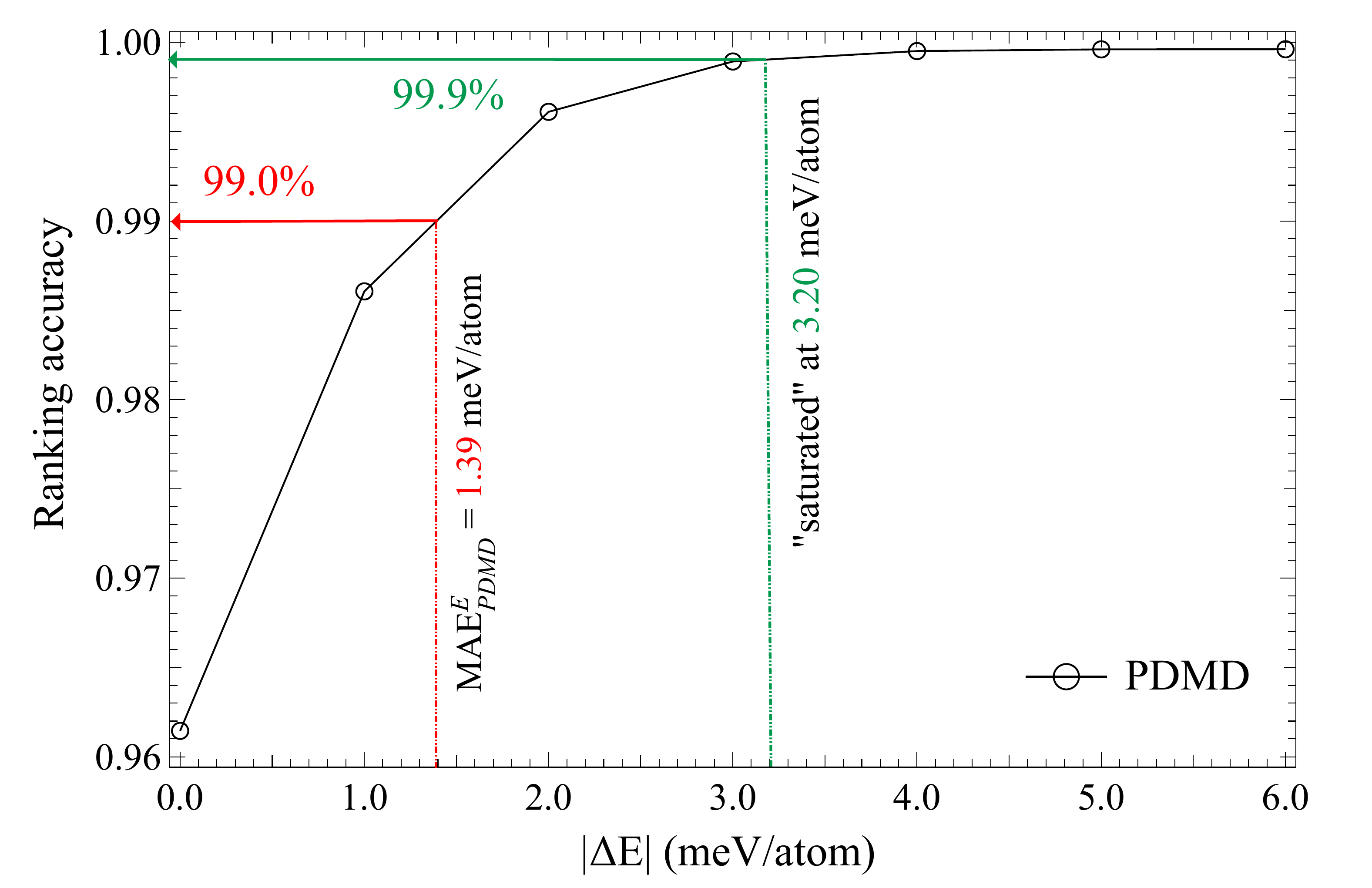}
\caption{Ranking accuracy of PDMD model with respect to DFT/PBE for system energy.}
\label{fig:rank}
\end{figure}

An iterative self-consistent approach, as detailed in the Methods section, was used to construct the PDMD model by minimizing the discrepancies in both system energy, $\textit{E}$, and atomic forces, ${\overrightarrow{F_i}}$, between PDMD and DFT/PBE for (H\textsubscript{2}O)\textsubscript{1$\leq$n$\leq$1000} clusters. As shown in Fig. \ref{fig:acc_ChemGNN}(a), our PDMD model achieves a mean absolute error (MAE) for $\textit{E}$ of 1.39 meV/atom. This energy MAE averaged over the water cluster size is significantly smaller than $k_BT$ at room temperature, the threshold to distinguish energy levels in defiance of thermal fluctuations. As for ${\overrightarrow{F_i}}$, a small cluster-size-averaged MAE of 50.7 meV/Å was achieved, leading to a remarkable coefficient of determination, $R\textsuperscript{2}=0.98$ (Fig. \ref{fig:acc_ChemGNN}(b)). As an additional validation, we randomly generated a benchmark set with 42,448 structures from 46 different (H\textsubscript{2}O)\textsubscript{n} clusters (i.e., 1,120 and 112 structures for each of the (H\textsubscript{2}O)\textsubscript{n$\in\{{1,2,...,30\}}\cup\{40,50,...,100\}$} and (H\textsubscript{2}O)\textsubscript{n$\in\{200,300,...,1000\}$}, respectively) by AIMD, and then formed all unordered structure pairs by taking every 2-element combination within each cluster size, yielding a total of 23,241,624 pairs (i.e., $\frac{37\times1120\times1119}{2}+\frac{9\times112\times111}{2}$). Then, we ranked the two structures in each pair by their energies using both PDMD and DFT/PBE. It turns out that our PDMD model achieves a 99.0\% rank-order agreement with DFT/PBE when the DFT/PBE energy difference, $|\Delta E|$, between the two structures is greater than 1.39 meV/atom (Fig. \ref{fig:rank}), the energy uncertainty of PDMD. If $|\Delta E|$ is raised to 3.20 meV/atom (i.e., $\sim\frac{1}{10}k_{B}T$ at 300 K per atom), the ranking accuracy is saturated at 99.9\%, making PDMD almost indistinguishable from DFT/PBE for MD simulations at room temperature.

\begin{figure}[!h]
\hspace{6pt}
\centering
\includegraphics[width=1.0\textwidth]{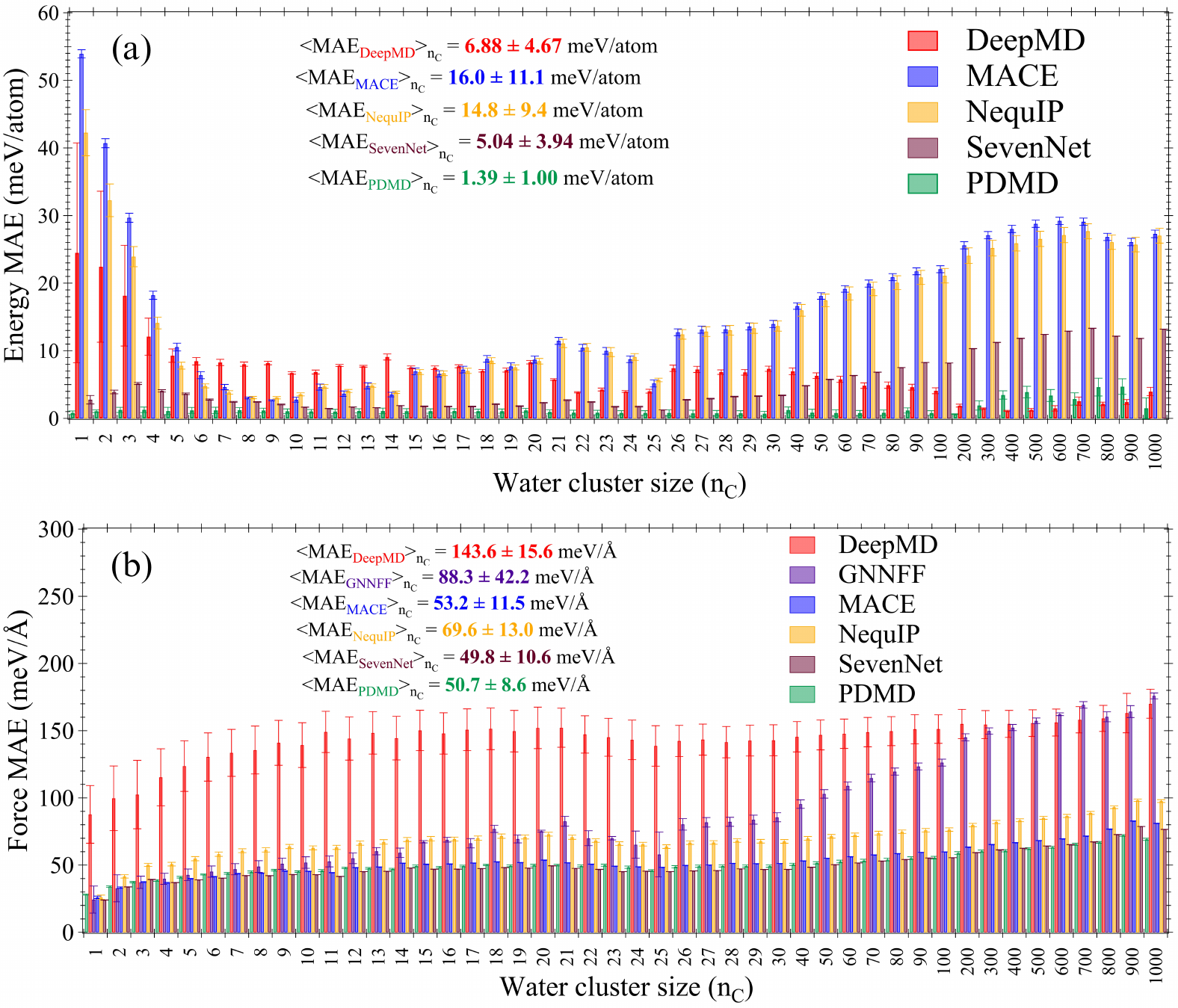}
\caption{Performance comparison of DeepMD, GNNFF, MACE, NequIP, SevenNet, and PDMD with respect to (a) system energy, \textit{E}, and (b) atomic force, ${\overrightarrow{F_i}}$, showing the average MAEs and their standard deviations over water cluster size, $n_{C}$. In addition, the error bars of \textit{E} and ${\overrightarrow{F_i}}$ for each water cluster size across three independent experiments are also presented.} 
\label{fig:acc_comp}
\end{figure}

Interestingly, the MAE of ${\overrightarrow{F_i}}$ obtained by PDMD at 50.7 meV/Å is significantly smaller than 140.0 meV/\AA, the \text{MAE} between two popular DFT functionals, namely  BLYP\cite{becke1988density} and PBE\cite{perdew1996generalized} (Fig. S1), further attesting to our PDMD model's quality, which is on par with $\textit{ab initio}$ theories. A more informative assessment of PDMD is to compare its performance with other molecular ML models' using the same dataset. In Fig. \ref{fig:acc_comp}(a), PDMD considerably outperforms DeepMD\cite{zeng2023deepmd}, MACE\cite{Batatia2022Design}, NequIP\cite{batzner20223}, and SevenNet\cite{park_scalable_2024} by at least 2.6x (i.e., 1.39 vs. 5.04 meV/atom) in energy MAE, especially for small clusters. For instance, PDMD's energy MAE stays at $\sim$1.0 meV/atom for (H\textsubscript{2}O)\textsubscript{1$\leq$n$\leq$100}, while its DeepMD counterpart drastically fluctuates from $\sim$5.0 to $\sim$25.0 meV/atom.\ Only for (H\textsubscript{2}O)\textsubscript{300$\leq$n$\leq$900}, PDMD is slightly worse than DeepMD, possibly due to the scarcity of training data. Nevertheless, PDMD is notably superior to DeepMD\cite{zeng2023deepmd} for force prediction by a large margin of $\sim$3x (i.e., 50.7 vs. 143.6 meV/Å as shown in Fig. \ref{fig:acc_comp}(b)) across the full spectrum of water cluster sizes. Although SevenNet significantly improved the average force MAE to 49.8 meV/\AA, which is comparable with PDMD, its average energy MAE of 5.04 meV/atom is still much worse than PDMD due to its energy MAE's rapid deterioration with increasing water cluster size, as indicated by a sizable standard deviation of 3.94 meV/atom (Fig. \ref{fig:acc_comp}(a)) that makes its applicability to larger systems questionable.  Taken together, our PDMD model offers a more practical data-driven solution for ultra-fast general-purpose MD simulations while retaining $\textit{ab initio}$ accuracy.  

\subsection*{b. Optimized Structures of Small Water Clusters, (H\textsubscript{2}O)\textsubscript{1$\leq$n$\leq$5}}
\begin{figure}[!ht]
\centering
\includegraphics[width=0.65\textwidth]{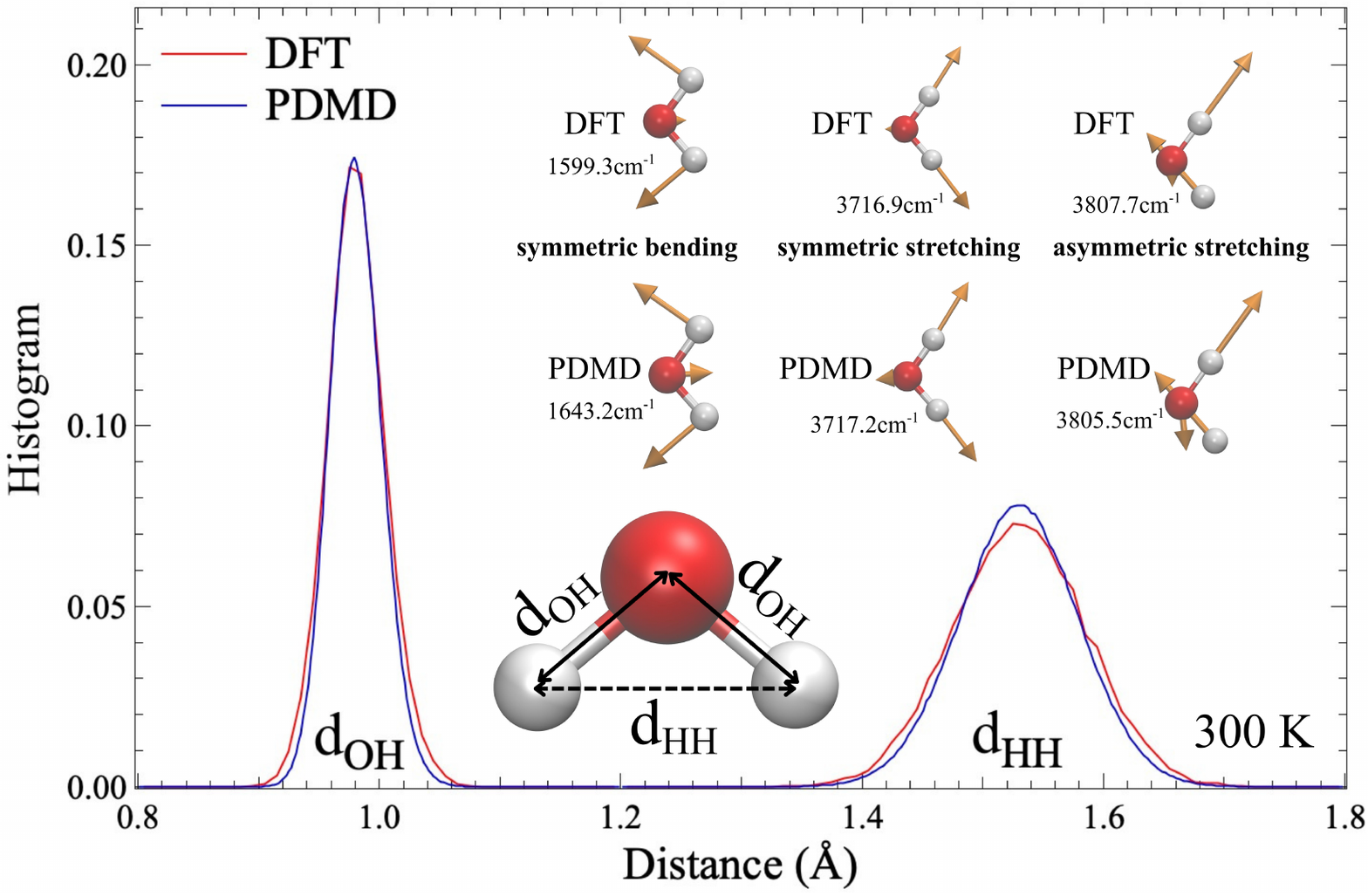}
\caption{PDMD-optimized structure of a water monomer and its O-H and H-H distance distributions in a PDMD simulation at 300 K, compared with DFT.}
\label{fig:monomer_struct}
\end{figure}

\begin{figure}[!htp]
\centering
\includegraphics[width=0.79\textwidth]{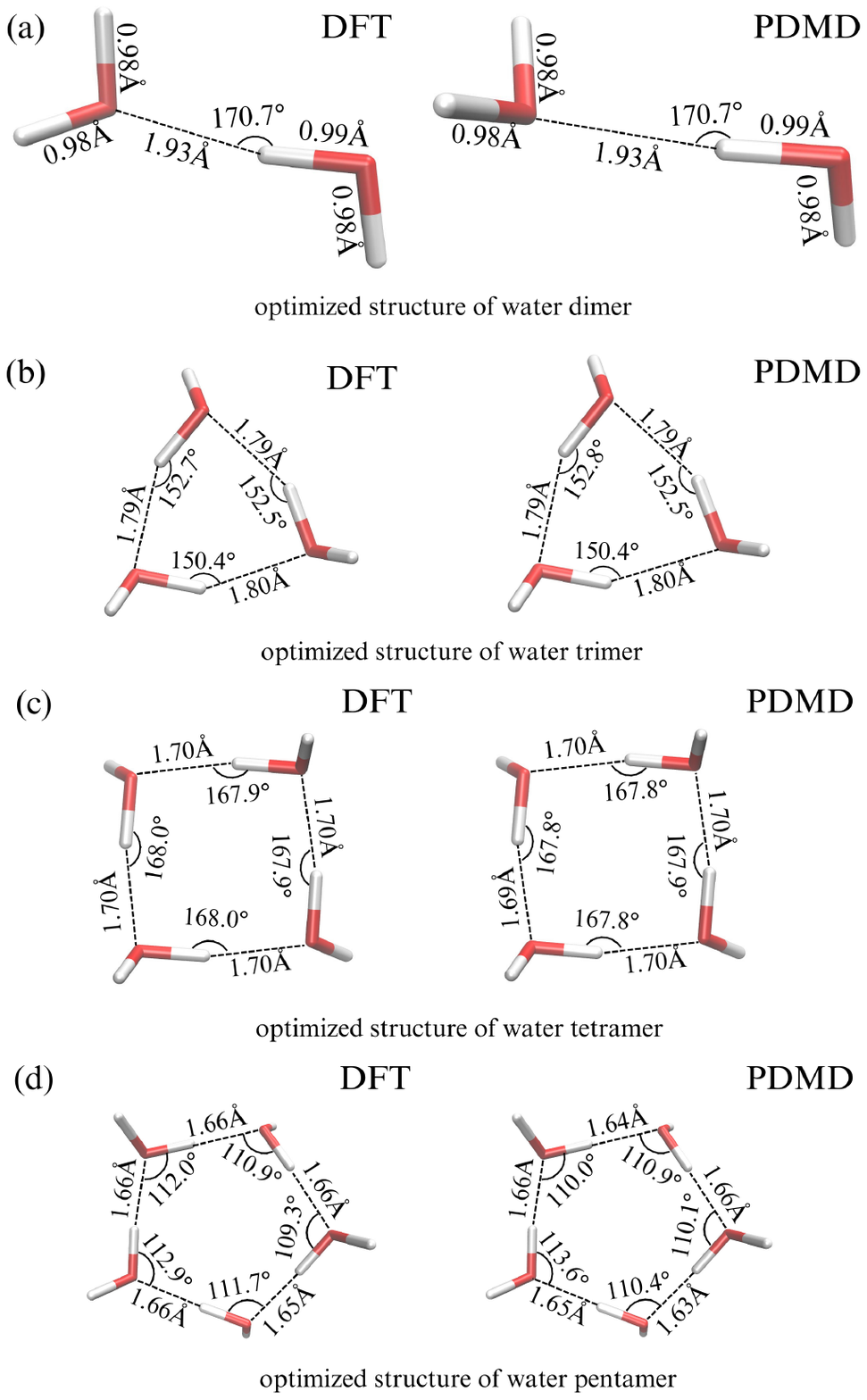}
\caption{DFT- and PDMD-optimized structures for water (a) dimer, (b) trimer, (c) tetramer, and (d) pentamer.}
\label{fig:water_struct}
\end{figure}

Geometry optimization of a chemical compound is driven by finite ${\overrightarrow{F_i}}$ until all forces vanish and the Hessian matrix is positive definite. Therefore, our PDMD model is expected to afford similar optimized structures compared to DFT/PBE if it can satisfactorily reproduce first- and second-order gradients of \textit{E}. Since the hexamer, (H\textsubscript{2}O)\textsubscript{6}, is the smallest water cluster that can populate multiple stable isomers at room temperature due to their nearly degenerate energy levels\cite{wang2012water}, a unique optimized structure exists only for each of the five smaller water clusters, i.e., (H\textsubscript{2}O)\textsubscript{1$\leq$n$\leq$5}. As shown in Fig. \ref{fig:monomer_struct}, the water monomer optimized by PDMD exhibits an O-H bond length of 0.978 \AA, the same as that optimized by DFT/PBE. Likewise, its H-H distance of 1.523 \AA ~is only 0.003 \AA ~shorter than its DFT/PBE counterpart. Interestingly, the monomer's distribution of O-H and H-H distances in PDMD at 300 K is in excellent agreement with AIMD, demonstrating reliable bond stiffness and angular rigidity. In addition, the water monomer's three vibrational modes ascertained by PDMD are similar to their DFT/PBE counterparts with two high-frequency stretching modes over 3700 cm\textsuperscript{-1} and one low-frequency bending mode near 1600 cm\textsuperscript{-1}. Unlike the water monomer, the optimized structures of all other clusters are stabilized by hydrogen bonds wherein a hydrogen atom covalently bonded to an oxygen atom is attracted towards another oxygen atom nearby through electrostatic interactions. As a result, an O-H bond that participates in hydrogen bonding is slightly longer than a non-hydrogen-bonded (``dangling") O-H bond because of partial electron transfer through the bridging hydrogen atom. This tiny yet critical bond elongation is precisely captured by both PDMD-optimized and DFT-optimized water dimers, which manifest a hydrogen-bonded O-H bond length of 0.99 \AA ~that is 0.01 \AA ~longer than its dangling counterpart (Fig. \ref{fig:water_struct}(a)). To maximize their hydrogen bonds\cite{malloum2019structures}, the water trimer, tetramer, and pentamer are known to form cyclic pseudo-planar optimized structures, which PDMD reproduces in close agreement with DFT in Fig. \ref{fig:water_struct}(b-d). In these three clusters, (H\textsubscript{2}O)\textsubscript{3$\leq$n$\leq$5}, the deviations between PDMD and DFT on hydrogen bond lengths and angles are less than 0.02 Å and 2.0\textdegree, respectively, further justifying the reliability of PDMD in modeling hydrogen bonds.   

\subsection*{c. Hydrogen Bonds in Large Water Clusters, (H\textsubscript{2}O)\textsubscript{n$\geq$6}}
\begin{figure}[!htp]
\centering
\includegraphics[width=0.60\textwidth]{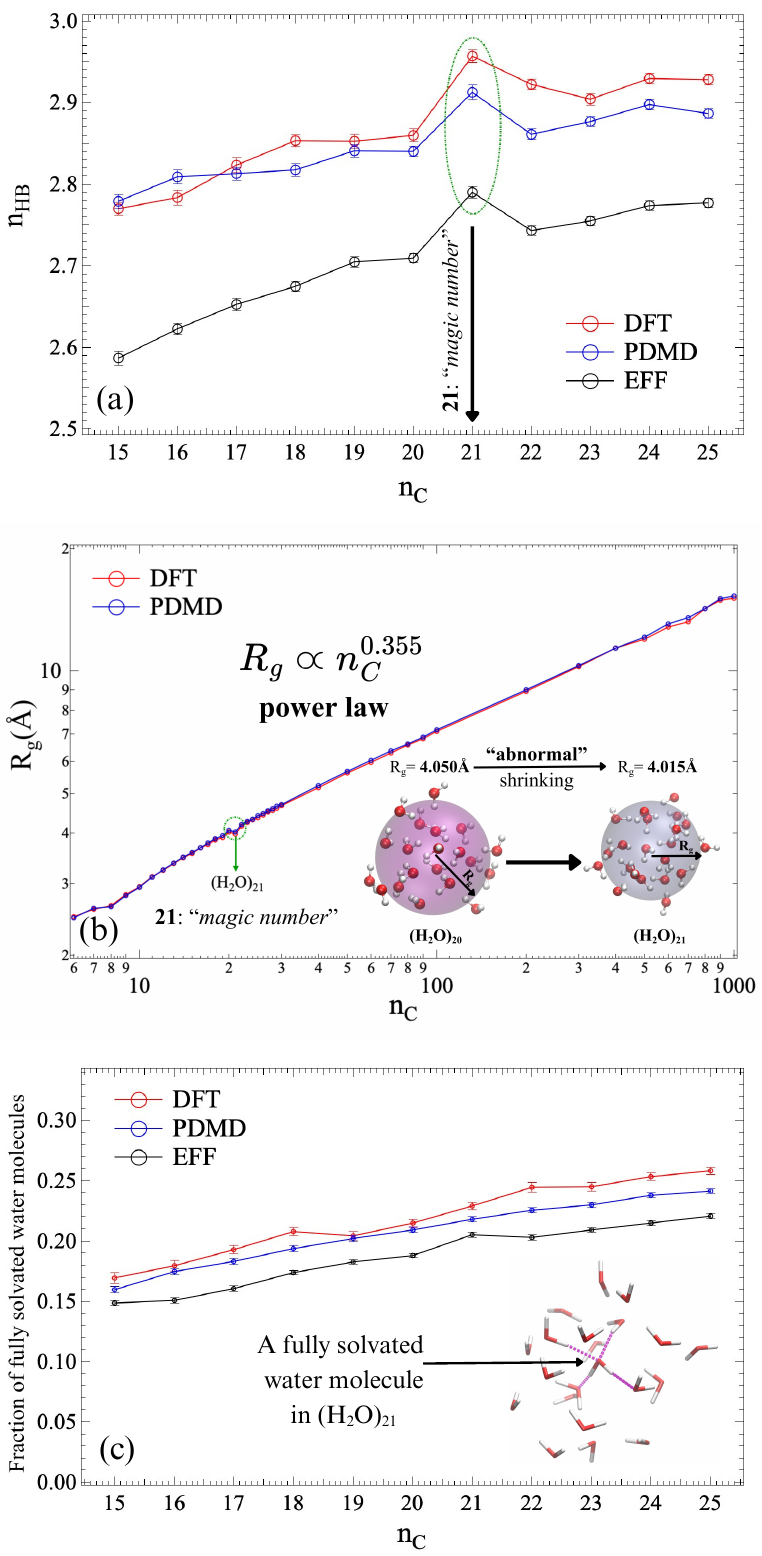}
\caption{ Dependence of (a) hydrogen bond number per water molecule, (b) radius of gyration, and (c) appearance probability of fully solvated water molecules on water cluster size. All results were extracted from MD trajectories at 300 K.}
\label{fig:hydrogen_bonds}
\end{figure}

\begin{figure}[!htp]
\centering
\includegraphics[width=0.60\textwidth]{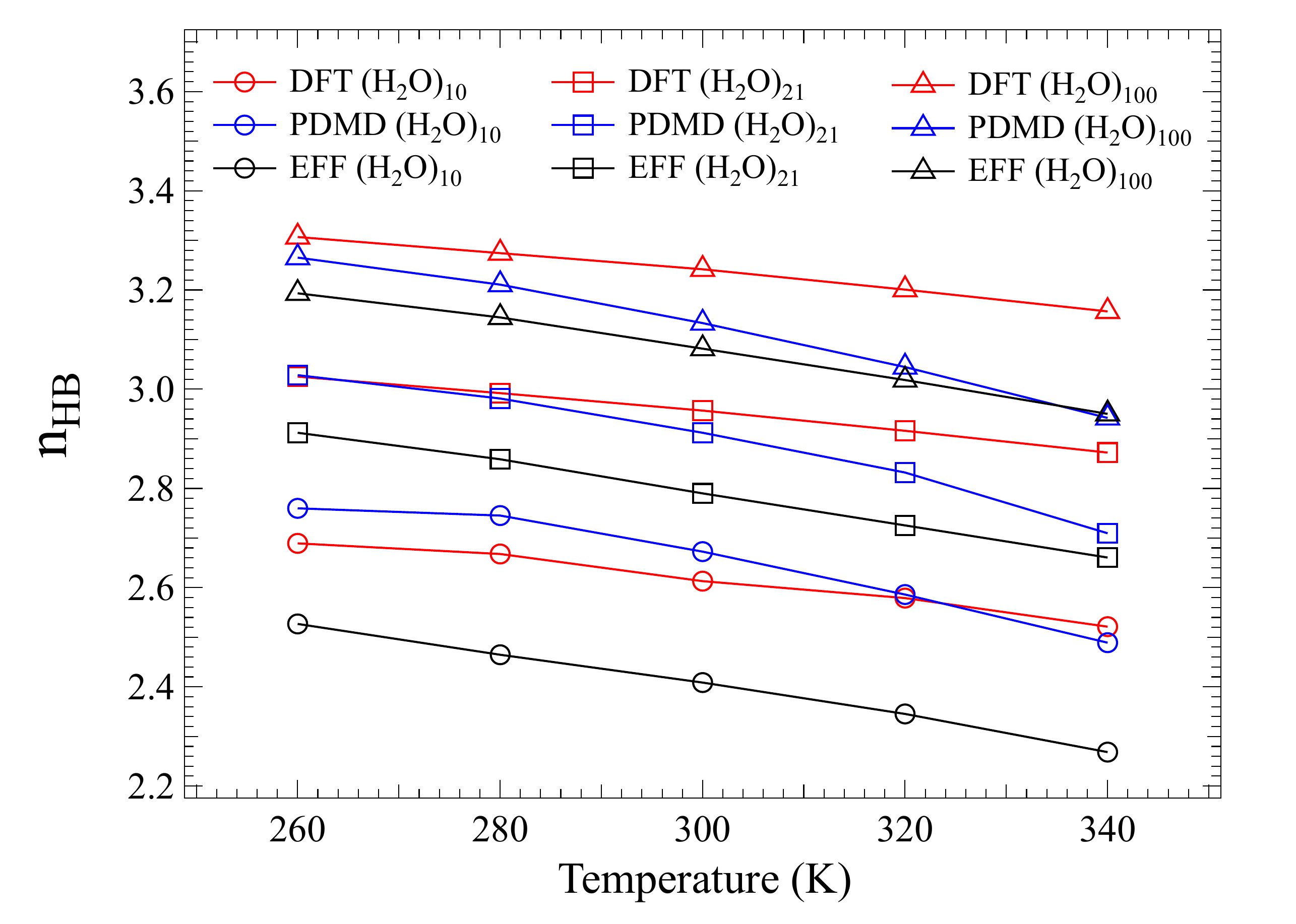}
\caption{ Temperature dependence of the number of hydrogen bonds per water molecule for (H\textsubscript{2}O)\textsubscript{10}, (H\textsubscript{2}O)\textsubscript{21}, and (H\textsubscript{2}O)\textsubscript{100} from 260 K to 340 K.}
\label{fig:temperature_effect_hydrogen_bonds}
\end{figure}

The substantially increased variety of hydrogen bonding topologies in large water clusters, (H\textsubscript{2}O)\textsubscript{n$\geq$6}, allows them to adopt non-planar versatile structures featuring facile hydrogen bonds that could be easily broken and formed at room temperature. Typically, a hydrogen bond is considered to appear when the distance between the donor and acceptor oxygen atoms is shorter than 3.5 \AA, and the angle between the donor, hydrogen, and acceptor atoms is greater than 120\textdegree. Using this criterion, we quantified how the number of hydrogen bonds per molecule, n\textsubscript{HB}, varies with cluster size, n\textsubscript{C}, in our PDMD trajectories. As shown in Fig. \ref{fig:hydrogen_bonds}(a), n\textsubscript{HB} generally grows monotonically with n\textsubscript{C} except at the "magic" number of 21, which affords a significantly larger n\textsubscript{HB} than its nearest neighbors on both sides. This abnormal spike in n\textsubscript{HB} has been ascribed to (H\textsubscript{2}O)\textsubscript{21}'s hydrogen-bonded network within a dodecahedral cage that creates a nearly spherical shell with minimal surface area, tightly binding all water molecules.\cite{ryding2015geometry}            Similarly, the trend was observed in AIMD and EFF trajectories (Fig. \ref{fig:hydrogen_bonds}(a)), although the latter significantly underestimates n\textsubscript{HB} compared to PDMD. Nevertheless, this apparent acceleration of hydrogen bond formation can be ascribed to the onset of a gas-to-liquid phase transition, accompanied by a seemingly aberrant cluster shrinking as cluster size increases through n\textsubscript{C}=21, as also illustrated in Fig. \ref{fig:hydrogen_bonds}(b). Under a pseudo-spherical approximation, the spatial dimension of a water cluster can be approximated by its radius of gyration, $R_g$, which is defined as $R_g=\sqrt{\frac{\sum_{i=1}^{N_s}\sum_{j=1}^{N_a}{|\overrightarrow{r_{ij}}-\overrightarrow{r_{i0}}|}^2}{N_sN_a}}$, where $N_s$ is the number of structure snapshots, $N_a$ is the number of atoms in a water cluster, and $\overrightarrow{r_{ij}}$ is the coordinate of the \textit{j}-th atom in the \textit{i}-th snapshot, whose center of mass is denoted as $\overrightarrow{r_{i0}}$. Although $R_g$ persistently increases with added water molecules up to n\textsubscript{C}=20 as anticipated, an unexpected reduction from $R_g$=4.050 Å to $R_g$=4.015 Å occurs at n\textsubscript{C}=21, wherein some water molecules start arranging themselves into tetrahedrally coordinated O-H chemical bonds and O-{}-H hydrogen bonds for a more compact spatial pattern as is always observed in liquid water. Our observation of this abnormal cluster shrinking is not only consistent with the AIMD simulations (inset of Fig. \ref{fig:hydrogen_bonds}(b)) but also agrees with the experimental finding that a minimum of 20 surrounding water molecules is needed to fully solvate one with two complete solvation shells evidenced by liquid water's ``fingerprint" band in (H\textsubscript{2}O)\textsubscript{21}'s infrared spectrum\cite{rognoni2021many}. Under a hard-sphere approximation, $R_g$ scales with the cluster size's cube root, i.e., $R_g\propto n_{C}^{1/3}$. Through a linear power-law fitting up to (H\textsubscript{2}O)\textsubscript{1000}, our PDMD yields an exponent of 0.355, in excellent agreement with the theoretical prediction if one accounts for water molecules' slight nonsphericity. Moreover, we calculated the appearance probability of fully solvated water molecules in our PDMD simulations (Fig. \ref{fig:hydrogen_bonds}(c)) for (H\textsubscript{2}O)\textsubscript{15$\leq$n$\leq$25}, and found a persistent rise that agrees with AIMD much better than EFF does, once again confirming PDMD's superiority over EFF for modeling hydrogen bonds. Interestingly, this superiority spans a wide temperature range, even though PDMD was developed using a training dataset generated only at 300 K. As shown in Fig. \ref{fig:temperature_effect_hydrogen_bonds}, PDMD's agreement with DFT on n\textsubscript{HB} is much better than that of EFF for (H\textsubscript{2}O)\textsubscript{10}, (H\textsubscript{2}O)\textsubscript{21}, and (H\textsubscript{2}O)\textsubscript{100} from 260 K to 340 K, spanning supercooled  to above physiological temperatures. As a result, PDMD can be readily deployed for temperature-varying simulations, such as parallel-tempering replica-exchange molecular dynamics\cite{sugita1999replica} for accelerated free-energy sampling and temperature-scheduled simulated annealing\cite{donnelly1987geometry} for geometry optimization.  

\subsection*{d. Many-Body Effects Exemplified by (H\textsubscript{2}O)\textsubscript{3$\leq$n$\leq$5}}
From the perspective of molecular coupling, the superior performance of PDMD over EFF, as illustrated by Fig. \ref{fig:hydrogen_bonds}(a) and Fig. \ref{fig:hydrogen_bonds}(c), stems from PDMD's intrinsic encoding of many-body interactions, which are completely missing in EFF. According to the many-body expansion (MBE) theory,\cite{richard2014understanding} the \textit{n}-th order approximation of a  system's energy, ${E^{n}}$, by accounting for all molecular interactions up to the \textit{n}-th order is given by ${E^{n}=\sum_{m=0}^{n-1}(-1)^mC_{m}^{n_c-n-1+m}\sum_{K=1}^{C_{n-m}^{n_c}}E_{K}^{(n-m)}}$, where $n_C$ is the number of monomers, ${C_{m}^{n}}$ is the binomial coefficient, and $E_{K}^{(n)}$ is the energy of the \textit{K}-th \textit{n}-mer. For example, ${E^{2}}$ of a water trimer, (H\textsubscript{2}O)\textsubscript{3}, is $E^{2}=\sum_{I=1}^{3}\sum_{J>I}^{3}E_{IJ}-\sum_{I=1}^{3}E_{I}$ if each water molecule is treated as a monomer, where $E_{IJ}$ is the energy of a water dimer, (H\textsubscript{2}O)\textsubscript{2}, consisting of the \textit{I}-th and \textit{J}-th water molecules, and $E_{I}$ is the \textit{I}-th water molecule's energy. As a result, the three-body contribution to energy, $E^{3B}$, in (H\textsubscript{2}O)\textsubscript{3} is given by $E^{3B}=E^{3}-E^{2}=E_{123}-E_{12}-E_{13}-E_{23}+E_{1}+E_{2}+E_{3}$. Generalized to arbitrary order, the many-body contribution, $E^{MB}$, beyond the two-body approximation in any molecular system with $N_C$ monomers reads $E^{MB}=E^{N_C}-E^{2}$. For the DFT-optimized (H\textsubscript{2}O)\textsubscript{3}, (H\textsubscript{2}O)\textsubscript{4} and (H\textsubscript{2}O)\textsubscript{5} as shown in Fig. \ref{fig:water_struct}, we calculated their $E^{MB}$ using both DFT and PDMD to assess the performance of PDMD on modeling the many-body effect, which is particularly critical for forming hydrogen bonds in water clusters. As listed in Table 1, PDMD effectively captures more than 92\% of DFT's many-body energy, making it a highly appealing linear-scaling approach to overcome the steep nonlinearity of many-body EFF, i.e., $O(n_C^{n})$, which becomes computationally intractable for $n>3$ even on systems with fewer than 100 water molecules. This salient capability arises from the chemical environment adaptive learning (CEAL) convolution, detailed in Methods (b), which makes intermolecular interactions depend on the surrounding local chemical environment. As a result, ChemGNN can effectively capture the many-body effects that govern water clusters.

\begin{table}
    \centering
    \Large
    \begin{tabular}{ccccc}
    \hline
         &(H\textsubscript{2}O)\textsubscript{n}&$E_{DFT}^{MB} (eV)$&$E_{PDMD}^{MB} (eV)$&$E_{PDMD}^{MB}/E_{DFT}^{MB}$ \\
         \hline
         &n=3  &-0.184  &-0.171 &92.9\%\\
         &n=4  &-0.468  &-0.447 &95.5\%\\
         &n=5  &-0.754  &-0.704 &93.4\%\\
    \hline
    \end{tabular}
    \caption{$E^{MB}$ in water trimer, tetramer, and pentamer calculated by DFT and PDMD.} 
    \label{tab:placeholder}
\end{table}

\subsection*{e. Linear Scaling of PDMD over System Size}
\begin{figure}[!htp]
\centering
\includegraphics[width=0.60\textwidth]{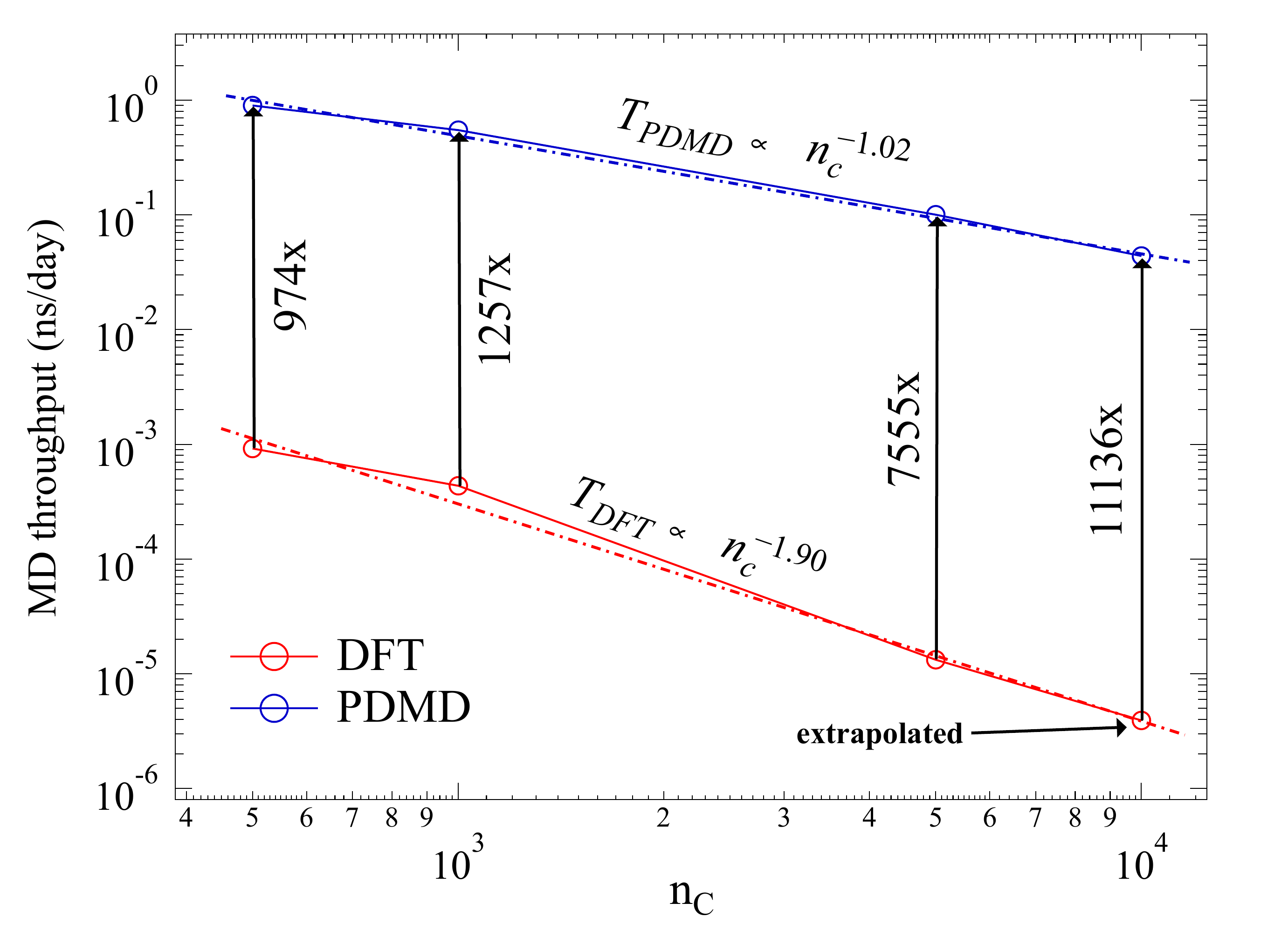}
\caption{ Molecular Dynamics throughput by DFT and PDMD as a function of the number of water molecules on a single Intel Xeon Platinum 8380 node.}
\label{fig:linear_scaling}
\end{figure}

Since our ultimate goal is to develop an ML approach to retain AIMD accuracy with orders-of-magnitude lower cost, the computational efficiency of PDMD was measured by the number of molecular dynamics steps delivered per day on a single Intel Xeon Platinum 8380 node using a timestep of 1.0 fs. As shown in Fig. \ref{fig:linear_scaling}, PDMD's throughput, $T_{PDMD}$, is inversely proportional to system size over a wide range from (H\textsubscript{2}O)\textsubscript{500} to (H\textsubscript{2}O)\textsubscript{10000}, i.e., $T_{PDMD}\propto n_C^{-1.02}$. By contrast, DFT's throughput, $T_{DFT}$, is approximately proportional to the inverse square of system size, i.e., $T_{DFT}\propto n_C^{-1.90}$, greatly limiting its applicability to large systems. Moreover, the memory demand of DFT also grows quadratically with system size, resulting in memory overflow on (H\textsubscript{2}O)\textsubscript{10000} even with 4 TB of memory on a single compute node. As a result, $T_{DFT}$ for $n_C=10,000$ in Fig. \ref{fig:linear_scaling} was extrapolated from the three other data points of $T_{DFT}$ due to our computer hardware limit. Nevertheless, the numerical performance gap between PDMD and DFT persistently widens from $\sim1,000$x to $\sim10,000$x when the number of water molecules increases from 500 to 10,000. This $\sim10^4$ difference in computational efficiency would enable PDMD to achieve results in minutes rather than the years required by DFT. It is worth noting that the computational efficiency of PDMD could be further improved by adopting a parallelization scheme that distributes its computational tasks across multiple compute nodes and/or acceleration devices, such as graphics processing units (GPUs).      

\section*{Discussion}

Over the past decade, the rapid advancement of artificial intelligence has inspired the development of machine learning molecular dynamics (MLMD),\cite{unke2021machine} which shows promise in greatly accelerating high-fidelity MD simulations, especially for systems with facile chemical bonds that demand accurate evaluation of system energy and atomic forces. For example, the hydrogen bond strength in water is within the kcal/mol range,\cite{silverstein2000strength} making the dynamic breaking and formation of hydrogen bonds frequent in aqueous solutions. Unlike other MLMD approaches that produce pairwise interatomic potentials,\cite{zeng2023deepmd,musaelian2023learning} our PDMD affords system energy and atomic forces directly from atomic coordinates and atom types, which are its sole inputs. This unique feature allows PDMD to capture the many-body effect efficaciously without being limited to the pairwise approximation, which breaks down in many aqueous systems due to the intrinsic non-additive interactions within them.\cite{herman2022classical} Although explicitly learned three-body potentials can partially encode this critical many-body effect as shown in a bcc tungsten crystal using cubic B-spline basis functions,\cite{xie2023ultra} the broken symmetry in amorphous systems such as liquid water would make the potential energy surface too complicated for any predefined basis functions due to the complexity of geometric features. Our PDMD model can address this challenge well by employing multiple aggregators with self-adaptive weights to accurately encode an atom's chemical environment for arbitrary atomic spatial patterns. As shown by our benchmark systems of (H\textsubscript{2}O)\textsubscript{1$\leq$n$\leq$1000} clusters, both gas-like and liquid-like molecular structures are satisfactorily delineated by PDMD, exhibiting a remarkably small energy MAE of 1.39 meV/atom, which is 2.6x better than its SevenNet counterpart (Fig. \ref{fig:acc_comp}(a)). Similarly, in terms of force MAE, PDMD outperforms DeepMD, GNNFF, and NequIP by $\sim$3x, $\sim$75\%, and $\sim$40\% (Fig. \ref{fig:acc_comp}(b)), respectively, while achieving the same level of accuracy as MACE and SevenNet. More importantly, the delicate topology of the hydrogen bond network is also precisely characterized by PDMD, leading to the consistent radius of gyration and the number of hydrogen bonds compared to AIMD for large water clusters (H\textsubscript{2}O)\textsubscript{n$\geq$6}. For small water clusters (H\textsubscript{2}O)\textsubscript{3$\leq$n$\leq$5},  their pseudo-planar optimized structures are well reproduced by PDMD. Most excitingly, PDMD also captured the abnormal cluster shrinking when the water cluster size approached 21, the experimentally observed magic number for the onset of the gas-to-liquid phase transition that can be ascribed to the accelerated nucleation with more fully solvated water molecules, each of which serves as the donor of two hydrogen bonds and the acceptor of two others.                                         

Given the high fidelity and exceptional efficiency of PDMD, chemical reactions in some large systems can be explored. One of them is the proton transfer in aqueous solutions. The energy barrier for proton transfer in liquid water is known to be as small as $\sim$1 kcal/mol, which is approximately the energy difference between a solvated Eigen and a solvated Zundel cation.\cite{swanson2007proton} More interestingly, the proton transfer proceeds through the Grotthuss mechanism, wherein a transferring proton is passed from one water molecule to another by breaking a hydrogen bond and forming another without requiring atomic displacement. This unique necessity of dynamic bonding topology invalidates the application of conventional empirical force fields with fixed chemical bonds. Therefore, proton transfer in aqueous solutions can only be treated by the computationally demanding AIMD or a slightly more cost-effective semi-empirical multi-state empirical valence bond (MS-EVB) theory\cite{wu2008improved}, which was parameterized to reproduce the quantum PES for protonated water dimer and tetramer. Nevertheless, MS-EVB's significantly underestimated proton diffusion coefficient\cite{wu2008improved} suggests an incomplete representation of the proton transfer reaction coordinate due to the missing solvation shell during its parameterization, which our data-driven PDMD can overcome through a sufficient sampling of fully solvated protonated water molecules under arbitrary pH. In fact, PDMD would be especially useful for low $[H^{+}]$ simulations, which demand a vast number of water molecules. Aqueous solutions containing anions are another class of systems that will benefit from PDMD. Anions are typically polarizable due to their rather diffuse electronic wavefunctions, making the many-body London dispersion interactions particularly prominent despite the strong screening effect in water. On that account, polarizable force fields\cite{warshel2007polarizable} have to be adopted to account for the mutual polarization of anions through a self-consistent approach, which notably compromises their numerical advantages over AIMD. By contrast, our potential-free PDMD naturally encodes the many-body interactions as much as is allowed by the statistical sampling of a chemical system with non-additive couplings, which are also critical in molecular solids\cite{jansen1962systematic} and biological systems\cite{ejtehadi2004three}. To address these challenges, we plan to refine our PDMD model by leveraging hypergraphs to capture the many-body nature of dispersion forces. In this approach, a hyperedge would connect all atoms involved in a particular collective interaction, thereby more accurately reflecting the many-body correlations than a simple pairwise graph. In summary, PDMD outperforms the current state-of-the-art approaches in machine learning molecular dynamics by offering chemically adaptive aggregators. It lays a solid foundation for future improvements through enhanced expressiveness and generalization of chemical hypergraphs\cite{konstantinova2001application}, particularly for chemical reactions in large systems, given PDMD's linear computational cost with respect to system size.

\section*{Methods}

\subsection*{a. Generation of Machine Learning Datasets} The construction of our ML datasets is based on an iterative self-consistent model training approach that will be delineated in Methods (c). Unless otherwise specified, all \textit{ab initio} calculations were performed by CP2K\cite{kuhne2020cp2k}, an open-source molecular simulation package, with the Goedecker–Teter–Hutter (GTH) pseudopotential\cite{goedecker1996separable}, Perdew–Burke–Ernzerhof (PBE) exchange-correlation functional\cite{perdew1996generalized}, polarized–valence–double-$\zeta$ (PVDZ) basis set\cite{woon1994gaussian}, and wavelet-based Poisson solver\cite{genovese2006efficient}. The initial structures of (H\textsubscript{2}O)\textsubscript{1$\leq$n$\leq$30} were obtained from a DFT study\cite{malloum2019structures} at the M06-2X/6-311++g(d,p) level of theory, while those of larger clusters, (H\textsubscript{2}O)\textsubscript{40$\leq$n$\leq$1000}, were generated by the Atomic Simulation Environment (ASE) package.\cite{hjorth2017atomic} Then, for each initial structure of (H\textsubscript{2}O)\textsubscript{1$\leq$n$\leq$21}, an AIMD simulation at 300 K was carried out to afford a 1-ns long trajectory before some snapshots were randomly selected. Subsequently, the \textit{E} and $\overrightarrow{F_i}$ of those chosen snapshots were evaluated to construct our ML dataset's first subset, which entails 63,000 (H\textsubscript{2}O)\textsubscript{n} structures. After the first round of machine learning model training, the optimized PDMD model was employed to generate a 200-ns long MLMD trajectory at 300 K, from which 100,000 more snapshots were randomly extracted to expand our ML dataset after their \textit{E} and $\overrightarrow{F_i}$ had been evaluated. For each subsequent round, 20,000 additional snapshots were extracted in the same manner. The expansion of this dataset only ceased when the desired accuracy of \textit{E} and $\overrightarrow{F_i}$ was achieved for our PDMD model. In total, our ML database consists of around 270,000 (H\textsubscript{2}O)\textsubscript{n} structures for (H\textsubscript{2}O)\textsubscript{1$\leq$n$\leq$21}. Afterwards, our pre-trained PDMD model for  (H\textsubscript{2}O)\textsubscript{1$\leq$n$\leq$21} was applied to each of the initial structures of  (H\textsubscript{2}O)\textsubscript{22$\leq$n$\leq$100} to generate a 1-ns long trajectory, from which a total of 35,840 structures were randomly selected to append to our ML database before we retrained the PDMD model to cover (H\textsubscript{2}O)\textsubscript{1$\leq$n$\leq$100}. Finally, by following the same protocol, 1,008 structures of (H\textsubscript{2}O)\textsubscript{200$\leq$n$\leq$1000} were added to our ML database to finalize training of our PDMD model. This increasing data scarcity with system size reflects the exponentially growing computational cost of DFT, which makes the ML dataset generation for systems with more than 1,000 water molecules numerically impractical, even with supercomputing facilities. Nevertheless, a separate benchmark database was assembled by extracting 1,120 and 112 structures for each of the (H\textsubscript{2}O)\textsubscript{n$\in\{{1,2,...,30\}}\cup\{40,50,...,100\}$} and (H\textsubscript{2}O)\textsubscript{n$\in\{200,300,...,1000\}$} water clusters, respectively, from their MD trajectories generated by our finalized PDMD model. Specifically, this benchmark database with 42,448 structures for (H\textsubscript{2}O)\textsubscript{1$\leq$n$\leq$1000} was used to assess the performance of DeepMD, GNNFF, MACE, NequIP, SevenNet, and PDMD for energy and force prediction against DFT, as shown in Fig. \ref{fig:acc_comp}.   

\subsection*{b. ChemGNN Model} The core of the ChemGNN model is the chemical environment adaptive learning (CEAL) convolutional layer (see Fig. \ref{fig:model_structure}), which is defined as follows.
\begin{gather}  
    \boldsymbol{h}_{i,A}^{(k)}= 
    \mathop{\oplus_{A}}_{j \in N(i)}\scalebox{0.8}{$MLP$}_{msg}^{(k)} \, (\boldsymbol{x}_{i}^{(k-1)}, \boldsymbol{x}_{j}^{(k-1)}, \boldsymbol{e}_{ji}), \; A\in \{0,...,M-1\} \\[3pt]    
    \boldsymbol{x}_{i}^{(k)}=\scalebox{0.8}{$MLP$}_{update}^{(k)} \, (\boldsymbol{x}_{i}^{(k-1)},w_{0}^{(k)}\!\cdot\! \boldsymbol{h}_{i,0}^{(k)},...,w_{M-1}^{(k)}\!\cdot\! \boldsymbol{h}_{i,M-1}^{(k)})
\end{gather} 

In the given equations, the variables $i$ and $j$ represent the central node and its adjacent nodes, denoted as $j\in N(i)$, respectively. $\boldsymbol{x}_{i}^{(k-1)}$ and $\boldsymbol{x}_{j}^{(k-1)}$ denote the node embeddings at the $(k\!\!-\!\!1)$-th convolutional layer for nodes $i$ and $j$. $\boldsymbol{e}_{ji}$ represents the edge embedding between nodes $j$ and $i$. Every node's embedding undergoes transformation by ${MLP}_{msg}^{(k)}$, consequently producing its respective message. Subsequently, these messages from the neighborhood are aggregated using the aggregator $\oplus_{A}$, where $M$ denotes the number of employed aggregators. In this work, we employ five aggregators: sum, mean, max, min, and standard deviation (std). In the model's $k$-th CEAL convolutional layer, we utilize trainable weights, represented as $\boldsymbol{w}_{A}^{(k)}$, to assign different levels of significance to the aggregated message $\boldsymbol{h}_{i, A}^{(k)}$. These weighted aggregated messages, $\boldsymbol{w}_{A}^{(k)}\!\cdot\! \boldsymbol{h}_{i, A}^{(k)}$, at the $k$-th layer, along with the central node's embedding from the previous $(k\!\!-\!\!1)$-th layer, are used to update the node's embedding through ${MLP}_{update}^{(k)}$.

\subsection*{c. PDMD Framework  
}
\begin{figure}[!tb]
\centering
\includegraphics[width=1\textwidth]{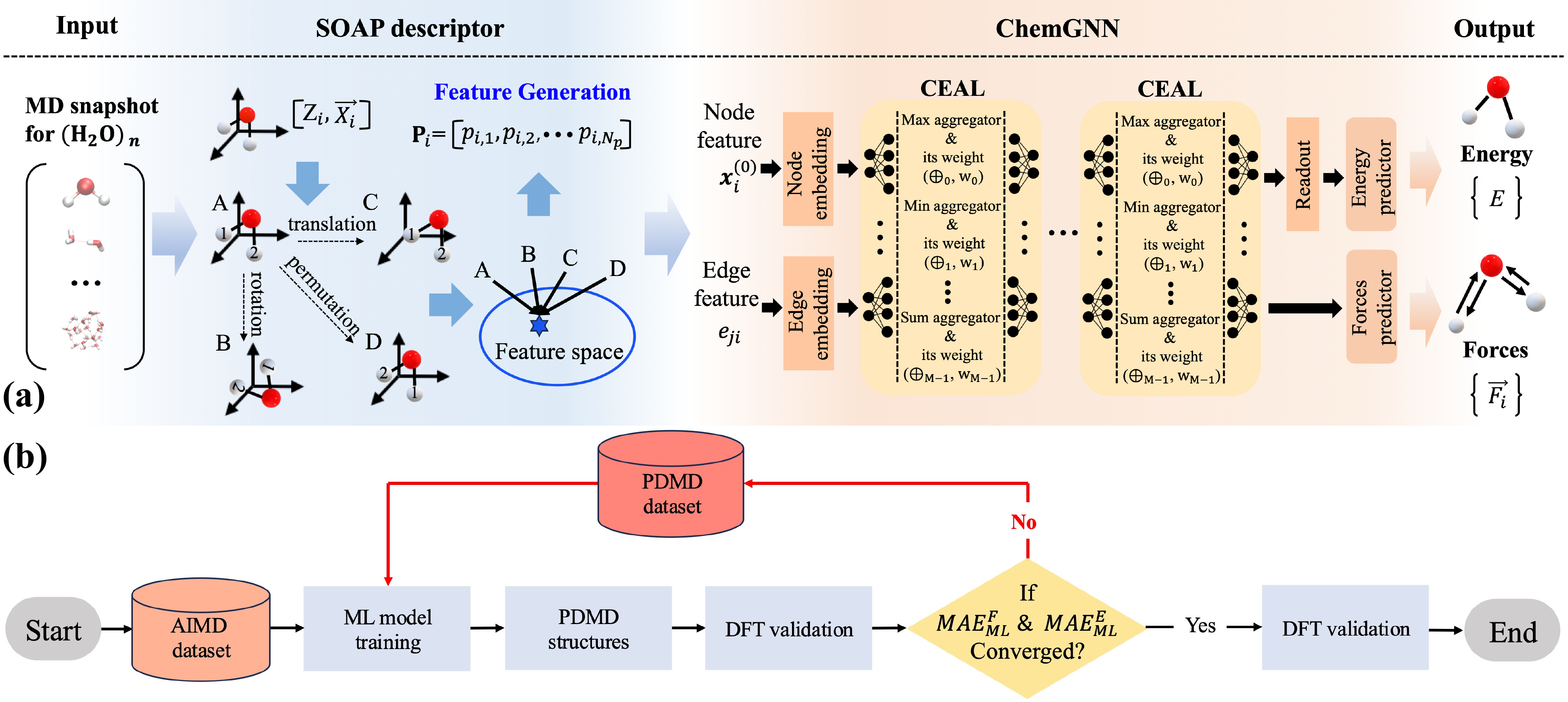}
\caption{(a) PDMD framework for arbitrary-sized water clusters (H\textsubscript{2}O)\textsubscript{$n$}, (b) PDMD self-consistent model training cycle. }
\label{fig:model_structure}
\end{figure}

The architecture of PDMD is illustrated in Fig. \ref{fig:model_structure}(a). It leverages the effective representations of atomic chemical environments and utilizes the ChemGNN model, which excels at adaptively capturing the features of local environmental information to predict the system's energy and atomic forces. 
The input dataset comprises snapshots of (H\textsubscript{2}O)\textsubscript{1$\leq$n$\leq$1000} obtained from MD simulations, including some configurations at the onset of the gas-liquid phase transition. Each snapshot provides the atom types $Z$\textsubscript{i} and Cartesian coordinates of atoms $\overrightarrow{X}$\textsubscript{i} as input. 
However, several challenges arise from the input representation. For example, the input's atomic coordinates depend on absolute position and orientation. Therefore, the Smooth Overlap of Atomic Positions (SOAP) descriptor\cite{de2016comparing} is used to enforce equivalence across equivalent structures. Specifically, SOAP utilizes orthogonal basis functions derived from spherical harmonics and radial basis functions to encode Gaussian-smeared atomic densities. By ensuring that all equivalent structures map to the same point in the feature space, SOAP further enriches the input atomic coordinates to invariant, higher-dimensional features, denoted as $\mathbf{p}_i=[p_{i,1}, p_{i,2},\dots, p_{i, N_p}]$, where $N_p$ represents the number of features for atom $i$.

Although the SOAP descriptor generates high-dimensional spatial representations, it does not include information about the atoms themselves. As a result, atom types are also incorporated as node features using a one-hot encoding vector, $\mathbf{o}_i$. Additionally, chemical couplings between atoms are represented as edges, with pairwise distances serving as edge features. In our model, we utilized distinct cutoff distances to determine bonds: 1.6 Å for H-H, 2.4 Å for H-O, and 2.8 Å for O-O. Each snapshot is thus represented as a graph $G = (V, E, D)$, where $V$ is the set of atoms with initial node features $\mathbf{x}_i^{(0)}=[\mathbf{o}_i, \mathbf{p}_i]$. $E$ represents the set of edges, $D$ is the adjacency matrix that encodes the molecular topology, with $D_{ji}=1$ indicating an edge (chemical bond) $e_{ji}$, and $D_{ji}=0$ otherwise. This graph representation is then processed using the ChemGNN model. Separate node and edge embedding layers are first applied to transform the node and edge features, enhancing their expressiveness. Next, the core CEAL layer of the ChemGNN model, equipped with an adaptive aggregation mechanism, is employed to fully extract essential information from various aspects of the nodes' chemical environments, thereby obtaining a high-level representation of the node. The model then splits into two pathways to predict the system's energy and atomic forces. In the first task, a readout layer performs pooling on the node embeddings to create a graph-level representation, which is passed through an Energy Predictor layer to predict the system energy $E$. In the second task, the node embeddings are directly fed into a Force Predictor layer to predict force vectors for each atom, $\overrightarrow{F_i}$.

PDMD enables both node- and graph-level predictions, making MLMD a practical tool for molecular dynamics. We can effortlessly generate numerous new structures through PDMD. Building on this capability, we employ an iterative process to expand the PDMD dataset by continuously generating and incorporating new snapshots, followed by retraining the model from scratch, as illustrated in Fig. \ref{fig:model_structure}(b). Initially, the AIMD dataset was the sole source for training the first version of the ML model. This model was then used for MD simulations to generate new snapshots, which were evaluated for model performance during this cycle. The newly generated PDMD data are added to the overall dataset for the subsequent training cycle. 
To determine model convergence, we monitor the difference in energy $E^\text{{MAE}}$ and forces $\overrightarrow{F}^\text{{MAE}}$ between cycle $l$ and cycle $l-1$. A convergence criterion is set, with energy and force differences limited to 1 meV/atom and 5 meV/\AA, respectively. The iteration stops automatically and the converged ML model is saved when both $\Delta {E}^\text{{MAE}} < 1$ meV/atom and $\Delta \overrightarrow{F}^\text{{MAE}} < 5$ meV/\AA. Our PDMD model achieved convergence after a few cycles.


\subsection*{d. Experimental Setup}
We utilized two sets of hyperparameters to train the models for predicting system energy and atomic forces, enabling PDMD to achieve near-optimal performance on both graph-level and node-level prediction tasks. Table S1 lists the hyperparameters employed to optimize the predictions of system energy and atomic forces. The hyperparameters remained unchanged in each cycle of model training refined by the self-consistent method. All reported MAEs in this work are the average of three independent experiments. For all PDMD simulations under a constant temperature, a timestep of 1.0 fs was adopted alongside a Nos\'e-Hoover thermostat.\cite{nose1984unified} The ASE package\cite{hjorth2017atomic} was chosen as our PDMD propagation engine, and the VESIN library\cite{metatensor-and-metatomic} was adopted as our neighbor list builder for an efficient graph construction in the ChemGNN architecture. As a result, in conjunction with the PyTorch Lightning framework, our PDMD model training efficiency scales remarkably well up to 32 NVIDIA Grace-Hopper GPU nodes before sub-linear roll-off, as shown in Fig. S2, enabling scalable artificial intelligence for massive AIMD datasets in the future.      

\section*{Data availability}
\large Our dataset for (H\textsubscript{2}O)\textsubscript{1$\leq$n$\leq$1000} water clusters is freely available at \url{https://taccchen.s3.us-east-2.amazonaws.com/PDMD_DATASET/PDMD_DATASET.tar.gz}. In addition, the same dataset in the formats compatible with DeepMD, MACE, NequIP, SevenNet, and GNNFF can also be downloaded from \url{https://taccchen.s3.us-east-2.amazonaws.com/PDMD_DATASET/PDMD_DATASET_Other_Formats.tar.gz}. 
\section*{Code availability}
\large Our PDMD code is freely available at \url{https://github.com/TACC/PDMD}. 

\section*{Acknowledgements}
Computational resources were partially provided by the Texas Advanced Computing Center at the University of Texas at Austin. We also wish to express our sincere gratitude to Ms. Qi Dai for the numerous insightful discussions.

\section*{Funding Statement}
Not applicable. 

\section*{Competing interests}
\large All authors declare that they have no known competing financial interests or non-financial interests that could have appeared to influence the work reported in this paper.

\section*{Author contributions}
H. C., M. C., and Y. W. conceived and supervised the project. H. Y. implemented the models and performed the experiments. H. C., M. C., and Y. W. performed the characterization and participated in the discussion. H. C., M. C., Y. W., and H. Y. wrote the original draft and created the visualizations. All authors read and agreed on the final version of the manuscript.

\section*{Additional information}
\textbf{Supplementary information} is available as a separate file. 

\bibliographystyle{naturemag}
\bibliography{references_abbreviated}

\end{document}